\documentclass[aps,pra,twocolumn,showpacs,superscriptaddress,superscriptaddress,amsthm,amsmath,amssymb]{revtex4-1}  
\usepackage[lofdepth,lotdepth]{subfig}
\usepackage{graphicx}
\usepackage{dcolumn}
\usepackage{bm}
\usepackage{multirow}
\usepackage{tikz}
\usepackage{pgfplots}
\pgfplotsset{compat=newest}

\usetikzlibrary{%
arrows,shapes.misc,shapes.arrows,matrix,%
positioning,scopes,decorations.pathmorphing,shadows,%
shapes.geometric,decorations.pathreplacing,shapes.gates.logic.US}

\tikzset{
m/.style={and gate US,text centered,minimum size=6mm,thick,draw=black,
top color=white,bottom color=white,font=\itshape},
nong/.style={rectangle,minimum size=6mm,thick,draw=white,
top color=white,bottom color=white,font=\itshape},
cont/.style={circle,text centered,inner sep=-1pt,minimum size=1mm,thick,draw=black,
top color=black,bottom color=black},
cxst/.style={circle,text centered,inner sep=0pt,minimum size=3mm,thick,draw=black,
top color=white,bottom color=white,append after command={
        [shorten >=\pgflinewidth, shorten <=\pgflinewidth, thick,]
        (\tikzlastnode.north) edge[thick] (\tikzlastnode.south)
        (\tikzlastnode.east) edge[thick] (\tikzlastnode.west)
        }},
g/.style={rectangle,minimum size=6mm,thick,draw=black,
top color=white,bottom color=white},
conn/.style={thick,draw=black,font=\ttfamily},
dconn/.style={thick,draw=black!30,font=\ttfamily},
conctrl/.style={thick,draw=black,font=\ttfamily},
>=latex,thick,
/pgf/every decoration/.style={/tikz/sharp corners},
point/.style={coordinate},
>=stealth',thick,draw=black,tip/.style={->,shorten >=0.007pt},
every join/.style={rounded corners},
hv path/.style={to path={-| (\tikztotarget)}},vh path/.style={to path={|- (\tikztotarget)}},
text height=1.5ex,text depth=.25ex}
{
  \tikzset{g/.append style={text height=1.5ex,text depth=.25ex}}
  \tikzset{nong/.append style={text height=1.5ex,text depth=.25ex}}
}

\pgfdeclarelayer{background}
\pgfdeclarelayer{foreground}
\pgfsetlayers{background,main,foreground}

\newcommand{\um}{\scalebox{0.5}[1.0]{\(-\)}}

\newcommand{\Had}[1][Had]{ \node (#1) [g]{$H$};}

\newcommand{\RY}[1][Rot]{ \node (#1) [g]{$R_y(\frac{\pi}{2})$};} 
\newcommand{\RYm}[1][Rot]{ \node (#1) [g]{$R_{\um y}(\frac{\pi}{2})$};}
\newcommand{\RX}[1][Rot]{ \node (#1) [g]{$R_x(\frac{\pi}{2})$};}
\newcommand{\RXm}[1][Rot]{ \node (#1) [g]{$R_{\um x}(\frac{\pi}{2})$};}
\newcommand{\RZ}[1][Rot]{ \node (#1) [g]{$R_z(\frac{3\pi}{2})$};}

\newcommand{\X}[1][1]{ \node (#1) [cxst] {};} 
\newcommand{\Z}[1][1]{ \node (#1) [cont] {};} 
\newcommand{\ctrl}[1][1]{ \node (#1) [cont] {};}

\newcommand{\mN}[1][mN]{ \node (w#1e) [point] {$X$};} 

\newcommand{\prO}[1][prO]{ \node (w#1b) [nong] {$|0\rangle$};}

\newcommand{\prPsi}[1][prPsi]{ \node (w#1b) [nong] {$|\psi\rangle$};} 
 
\newcommand{\wP}[1][wp]{\node(wp#1)[point]{};}
\newcommand{\wire}[1]{\draw[conn] (w#1b) -- (w#1e);}
\newcommand{\dline}[1]{\draw[dconn] (wpS#1) -- (wpF#1);}
\newcommand{\ctrlwire}[1]{\draw[conctrl] (S#1) -- (F#1);}

\begin{document}

\title{Optimization of Clifford Circuits}
\author{Vadym Kliuchnikov}
\email{v.kliuchnikov@gmail.com}
\altaffiliation{David R. Cheriton School of Computer Science, University of Waterloo, Waterloo, ON, Canada}
\affiliation{Institute for Quantum Computing, University of Waterloo, Waterloo, ON, Canada}
\author{Dmitri Maslov}
\email{dmitri.maslov@gmail.com}
\affiliation{National Science Foundation, Arlington, VA, USA}

\begin{abstract}
We study synthesis of optimal Clifford circuits, and apply the results to peephole optimization of quantum circuits.  We report optimal circuits for all Clifford operations with up to four inputs.  We perform peephole optimization of Clifford circuits with up to 40 inputs found in the literature, and demonstrate the reduction in the number of gates by about 50\%.  We extend our methods to the synthesis of optimal linear reversible circuits, partially specified Clifford unitaries, and optimal Clifford circuits with five inputs up to input/output permutation.  The results find their application in randomized benchmarking protocols, quantum error correction, and quantum circuit optimization.
\end{abstract}


\maketitle

\section{Introduction}
Randomized benchmarking protocols \cite{arXiv:0707.0963} are a promising approach to the experimental assessment and evaluation of quantum information processing proposals.  Experiments implementing these protocols were already demonstrated by multiple research groups \cite{rbmg, rb2q, arXiv:0808.3973}.  The advantages over other methods include the independence from the physical implementation details of those quantum information processing systems being tested \cite{rbmg, arXiv:0808.3973}, and scalability.  A randomized benchmarking protocol may be described as a repeated application of a set of randomly chosen Clifford operations, followed by the measurement.  Access to time optimal implementation of Clifford operations allows us to reduce the time required to perform a given benchmarking experiment, and thus it is important for present practical purposes. 

A goal of an experimentalist desiring to employ a randomized benchmarking protocol is to construct a complete set of physically implementable operations that can be used to generate any Clifford operation, and then be able to express any Clifford operation using the set of such implementable operations available.  Those implementable operations are furthermore referred to as elementary operations.  To illustrate, in \cite{rbmg} the set of elementary operations consists of the two-qubit phase gate (controlled-Z) and all single qubit Clifford and Pauli gates.  In \cite{arXiv:0808.3973}, the two-qubit ZZ-interactions are provided by the driving Hamiltonian, single qubit gates in the X-Y plane are implemented as RF pulses, and single qubit gates in the Z-plane are implemented through a frame change, and require no physical action (as such, they are ``free of charge'').  The amount of physical resources required to implement each elementary gate, as well as the very set of gates that may be implemented directly, 
varies from one quantum information processing proposal to another.  Unable to capture all possible elementary gate libraries and circuit cost metrics, we concentrated on the study of quantum circuits composed with Hadamard gate, Phase gate (and its inverse), and the two-qubit CNOT gate, and two simple metrics of circuit cost---the gate count and the depth.  However, we designed our algorithms and implementation such that they may be modified to accommodate essentially any gate library, as well as more sophisticated metrics of the circuit cost. 

In particular, we study the problem of the optimal synthesis of Clifford operations acting on a small number of qubits.  We determine the cost of the overall Clifford operation based on the number of single and two-qubit elementary operations required to implement it.  This constitutes a simple measure for estimating the difficulty of implementing Clifford operations in an experiment.  We synthesize optimal Clifford circuits acting on two to four qubits, and optimal Clifford circuits acting on five qubits and up to input/output permutation.  We use the optimal implementations of Clifford operations acting on a small number of qubits in peephole optimization \cite{pho} (a detailed description of peephole optimization, as used in this paper, may be found in Section \ref{subsec:peephole}) of larger Clifford circuits.  The experiments reveal substantial practical improvement in large-scale designs of Clifford circuits.  Finally, we apply the ideas developed in the paper to find an optimal encoding circuit for 
the $[[5,1,3]]$ five-qubit error correcting code.  This method can be applied to synthesize encoding circuits for other 
error correcting codes that use 
a small number of qubits.

Clifford circuits (also known as stabilizer circuits) have been studied well in the relevant literature:  \cite{stab} reports a decomposition of the $n$-qubit Clifford operations using at most $O(n^2/\log_2{n})$ gates; \cite{quant-ph/0703211} develops linear depth implementations.  Both papers report asymptotically optimal implementations, however, suboptimal in the absolute sense.  As reported in \cite{rbmg}, finding optimal implementations of Clifford circuits with up to two qubits is straightforward.  In our paper, we report optimal Clifford circuits for up to four qubits, optimal Clifford circuits up to input/output permutation for up to five qubits, and optimize scalable implementations of the Clifford circuits by a factor of roughly two.

\section{Preliminaries}
We assume reader's familiarity with the basic concepts in quantum computing, stabilizer formalism and Clifford circuits, and provide a very short introductory overview to remind basic concepts and introduce the notations.  For more information, please, refer to \cite{stab}.

Clifford quantum circuits consist of Hadamard ($H$), Phase ($P$, also known as $S$ gate) and CNOT gates.  The important property of these gates is that they map Pauli matrices
\[
X=\left(\begin{array}{cc}
0 & 1\\
1 & 0
\end{array}\right)\!,\: Y=\left(\begin{array}{cc}
0 & -i\\
i & 0
\end{array}\right)\!,\: Z=\left(\begin{array}{cc}
1 & 0\\
0 & -1
\end{array}\right) 
\]
and their tensor products into themselves by conjugation. In particular: 
\begin{align*}
HXH^{\dagger}&=Z,\:HYH^{\dagger}=-Y,\:HZH^{\dagger}=X, \\
PXP^{\dagger}&=Y,\:PYP^{\dagger}=-X,\:PZP^{\dagger}=Z. 
\end{align*}
The CNOT gate acts on two qubits and transforms Pauli matrices by conjugation as follows:
\begin{align*}
X\otimes I & \mapsto X\otimes X,\: Z\otimes I\mapsto Z\otimes I, \\
I\otimes X & \mapsto I\otimes X,\: I\otimes Z\mapsto Z\otimes Z.
\end{align*}

Compact representation of any unitary that can be computed by a Clifford circuit is a direct consequence of the Clifford gates' property described above. Action of a circuit on any input is uniquely defined by this representation~\cite{hrepr}. Taking into account the identity $Y=iXZ$, it suffices to know the action by conjugation of the $n$-qubit circuit on $2n$ Pauli matrices.  The result of the application of the circuit to each Pauli matrix can be encoded using $2n+1$ bits \cite{stab}. Each single-qubit Pauli matrix can be encoded in two bits, as follows: 
\[
 I\sim\left(0|0\right),X\sim\left(1|0\right),\: Z\sim\left(0|1\right)\; Y\sim\left(1|1\right).
\]
It is convenient to separate $X$ and $Z$ parts when encoding larger circuits: 
\[
I\sim\left(\mathbf{0}|\mathbf{0}\right), X\sim\left(1|0\right),-I\otimes X\sim\left(\mathbf{0}1|\mathbf{0}0|1\right).
\]
One additional bit shown at the end is used to encode the overall overall phase, here restricted to $\pm 1$. For any unitary the sign can be adjusted by applying the round of Pauli gates at the end of the computation. In most of our applications this can be done for free. As a result, we will consider only the $2n\times2n$ part of the encoding matrix. Commutativity, the relations between Pauli matrices are preserved under conjugation and induce additional constraint on the encoding matrix---it must be symplectic (a square block matrix $M=\left(\begin{array}{cc}
A & B\\
C & D
\end{array}\right)$ is called symplectic iff the following three conditions hold: $A^TC=C^TA$, $B^TD=D^TB$, and $A^TD-C^TB=I$).  Furthermore, the canonical decomposition theorem \cite{stab} shows that any binary symplectic matrix encodes some Clifford circuit. 

The above matrix representation can be efficiently updated \cite{stab} when adding new gates to the end of an existing circuit. Computationally, adding a gate requires updating one or two columns of the encoding matrix. In particular, the application of the Phase gate to qubit $k$ corresponds to the addition modulo 2 of column $k$ to column $n+k$, the Hadamard gate on qubit $k$ corresponds to exchanging columns $k$ and $n+k$, and the CNOT gate with control $k$ and target $j$ corresponds to the addition of column $k$ to column $j$ and the addition of column $n+j$ to column $n+k$. An empty Clifford circuit corresponds to the identity matrix.  These rules suffice to determine the $2n\times2n$ binary symplectic matrix encoding the unitary computed by a given Clifford circuit. 

For linear reversible circuits---those composed only with CNOT gates---it suffices to store only the top left $n\times n$ part of the binary symplectic matrix. The described procedure for updating columns immediately implies that the binary symplectic matrix for linear reversible circuits should be of the following form: 
\[
\left(\begin{array}{c|c}
A & 0\\ \hline 0 & B
\end{array}\right).
\]
As the matrix must be symplectic, we have $A^{T}B=I$ (per third equation in the definition), which uniquely determines $B$ given $A$.  Therefore, we can store linear reversible unitaries more efficiently than a generic Clifford operation.


The two optimality measures that we consider are the minimal number of the Clifford gates required and the minimal depth of the circuit implementing the given unitary.  For brevity, we call them the \emph{gate count} and the \emph{depth} of the unitary.  Our ideas extend to other optimality measures, such as the weighted gate counts/depth (including a popular CNOT-depth metric).

\section{Algorithms}
The main challenge in our approach to finding optimal circuits is coping with the size of the search space, that grows very rapidly, as illustrated in Table~\ref{tab:sz}.  The size of the database for Clifford unitaries acting on $n$ qubits reported in this table is calculated using the following formula: $4n^2 \times N_G / n!,$ where $4n^2$ corresponds to the storage space (in bits) required by a single unitary, $N_G$ is the number of elements in the respective Clifford group, and the division by $n!$ corresponds to the estimated reduction due to the input/output renaming invariance ($n!$ marks a maximal possible reduction, therefore the entire figure provides a lower bound on the size of the database).

Our algorithm is based on the Breadth First Search.  The number of distinct unitaries computed by Clifford circuits grows as $2^{\Theta(n^2)}$. We address the resulting challenge in several ways. First, each node of the search tree corresponds to an equivalence class of unitaries instead of the unitary itself.  Second, we use meet in the middle technique to avoid building the full tree~\cite{fbit}.  Finally, we use a special data structure to store the search tree in a compact way. It is described in more details in Section \ref{subsec:impl}.

\renewcommand{\arraystretch}{1.125}
\begin{table}
\begin{center}
\begin{tabular}{|c|c|c|c|}
\hline 
$G$ & $n$ & $N_G$ & $Size_{Gr} (GB)$ \tabularnewline \hline 
\hline 
\multirow{3}{*}{$Sp$} 
& 3 & 1,451,520 & $1.01 \times 10^{-3}$ \tabularnewline \cline{2-4}
& 4 & 47,377,612,800 & $14.71$ \tabularnewline \cline{2-4}
& 5 & 24,815,256,521,932,800 & $2.41 \times 10^6$ \tabularnewline \cline{2-4}
\hline 
\multirow{2}{*}{$Gl$} 
& 6 & 20,158,709,760 & $0.12$ \tabularnewline \cline{2-4}
& 7 & 163,849,992,929,280 & $185.44$ \tabularnewline \cline{2-4}
\hline 
\end{tabular}
\end{center}
\caption[Size of the symplectic part of Clifford group]
{\label{tab:sz}
$G$ -- group: $Sp$ -- symplectic part of Clifford group, $Gl$ -- group generated by linear reversible circuits;
$n$ -- number of qubits, 
$N_G$ -- size of the corresponding group,
$Size_{Gr} $ -- lower bound on the size of the database taking into account input/output renaming (GB).}
\end{table}

The equivalence relation we use to reduce the size of the search space is the following: two unitaries are equivalent if they can be computed by circuits that are the same up to simultaneous renaming of their inputs and outputs.  Both gate count and depth of a unitary are invariant with respect to such simultaneous renaming.  During the search we store only a canonical representative of each class. For $n$ inputs this results in a reduction of the number of unitaries to be stored by a factor of approximately $n!$.  The number of unitaries corresponding to the same canonical representative is not always $n!$, but this is the most common case. In particular, the fraction of four-qubit unitaries that have less than 24 ($=4!$) elements in their equivalence class is less than $9.7\times10^{-5}$. To search for five-qubit optimal Clifford circuits we used the equivalence relation corresponding to the independent renaming of the inputs and outputs, in other words, we ignored SWAP gates. This further shrinks the 
search space, but the results are suboptimal in the scenario when SWAP has a non-zero cost. 

The idea of the meet in the middle (MiM) technique is based on the optimality of subcircuits of any optimal circuit. Given a database $DB_c$ of all unitaries with the cost at most $c$, MiM allows us to find optimal circuits for unitaries with the cost at most $2c$. Suppose we are looking for an optimal circuit computing a unitary $f$ with cost $c+d\le2c$. We can always split the optimal circuit into two optimal circuits with $d$ and $c$ gates. Therefore, there always exist a unitary $g$ with cost $d\le c$ such that its composition with $f$ has cost $c$, and it is in our database. We can find $g$ by trying all unitaries from the database and checking if $g\circ f$ is also in the database. In the worst case, using meet in the middle increases the time required to find a circuit by a factor proportional to the size of $DB_c$, in comparison to using the database $DB_{2c}$. At the same time, meet in the middle significantly reduces the required memory. For example, in the case of four qubits the maximal number of 
gates required is 17 and the size of the database is 14.72 GB.  Using the database with optimal circuits up to 9 gates reduces the required memory to just 108 MB.  Meet in the middle is vital for the search of optimal five-qubit Clifford circuits up to input/output permutation. In this case, the size of the full database would have been about $2.41\times 10^6$ GB.  

\begin{figure*}[t]
\begin{center}
\includegraphics{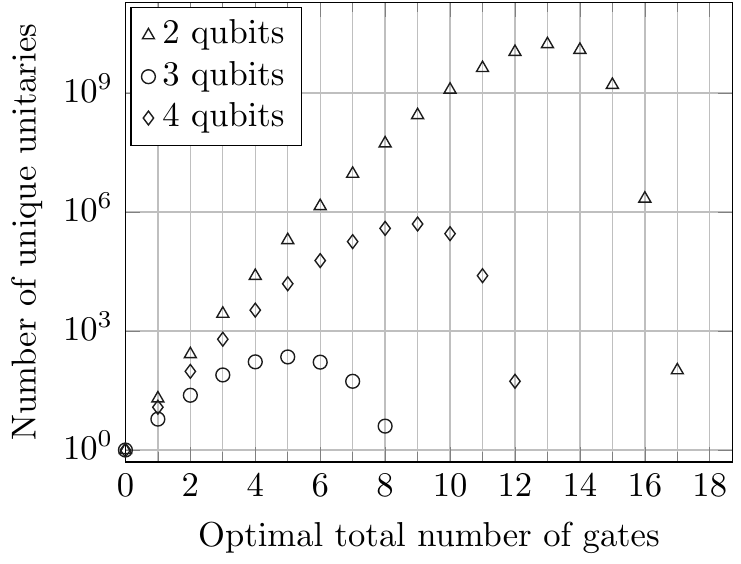}
\hspace{1.5cm}
\includegraphics{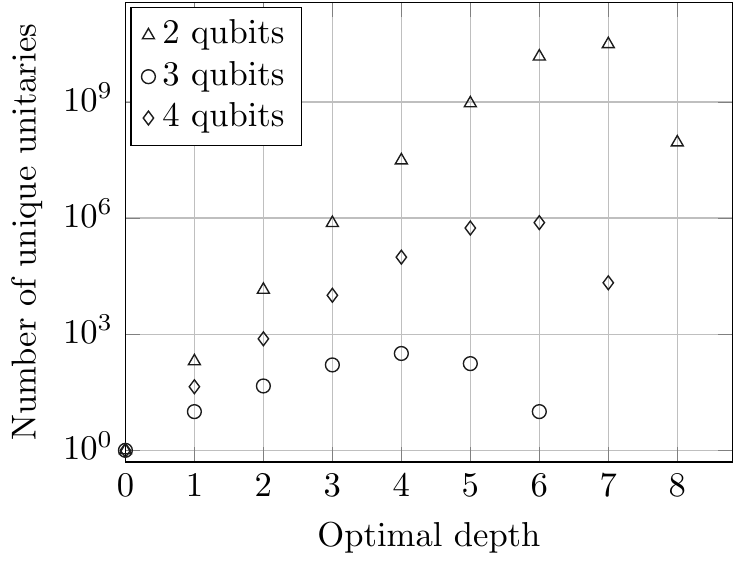}
\caption{\label{fig:cl24}The number of unique Clifford unitaries on 2, 3, and 4 qubits per optimal gate count and depth.}
\end{center}
\end{figure*}

\begin{figure}[t]
\begin{center}
\includegraphics{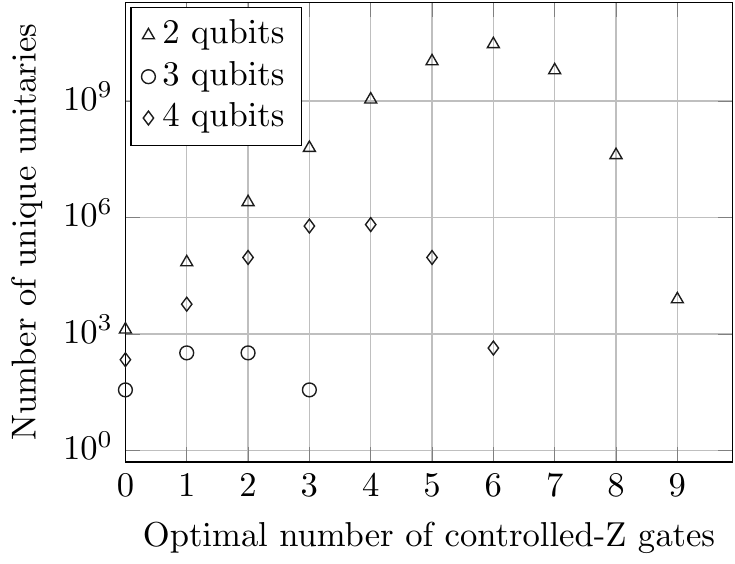}
\caption{\label{fig:cl24-cz}The number of unique Clifford unitaries on 2, 3, and 4 qubits per optimal number of controlled-Z gates.}
\end{center}
\end{figure}

\begin{figure}[t]
\begin{center}
\includegraphics{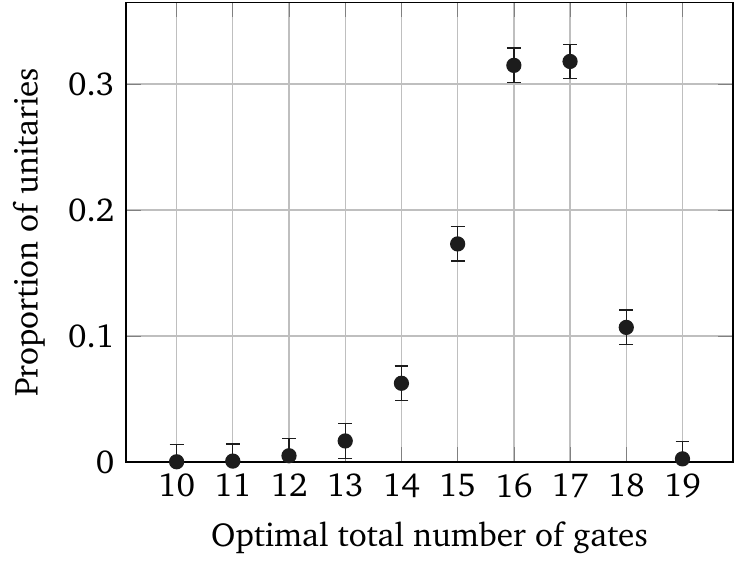}
\caption{\label{fig:cl5}Estimated proportion of the 5-qubit Clifford unitaries per optimal gate count (independent input/output renaming allowed).}
\end{center}
\end{figure}

\subsection{Computing canonical representative}
To find the canonical representative with respect to the {\em simultaneous} renaming of the inputs and outputs we compute all elements of the equivalence class, encode them as bit strings, and find the minimum.  We need to go though all possible permutations. This is accomplished by applying a single transposition at each step.  Exchanging inputs $k$ and $j$ of an $n$-qubit Clifford circuit corresponds to swapping columns and rows of the binary symplectic matrix. The pair of columns $(k,k+n)$ must be swapped with $(j,j+n)$, pairs of rows with the same indexes must be swapped also. Internally we represent each binary matrix as an array of integers. Each integer corresponds to a column of the binary symplectic matrix.  We precompute the required transpositions of the bit strings of length $2n$ and use a lookup table to speed up the swapping of rows of the binary symplectic matrix. 

When we allow the {\em independent} renaming of the inputs and outputs we apply a more efficient procedure for canonical representative computation. In most cases we have $(n!)^2$ representatives corresponding to the same equivalence class. First we find all $n!$ representatives corresponding to the different row permutations \footnote{By row permutation we mean a permutation acting simultaneously on first $n$ rows and rows $n+1,\ldots, 2n$. }. Then we store columns $k$ and $k+n$ together in one bit string and sort the resulting bit strings using a sorting network. This gives a canonical representative with respect to column permutation for a fixed row permutation. Finally, we encode the representative for each row permutation as a bit string and find the minimum.  

We apply the same idea for linear reversible circuits.  To exchange two inputs $k$ and $j$ we just need to swap columns $k$ and $j$ and rows $k$ and $j$ of the matrix encoding the circuit. This approach also extends to partially specified matrices. 

\subsection{Implementation details} \label{subsec:impl}
The main bottleneck in our search is the amount of memory available.  In addition to using canonic representation, we tried to minimize the memory overhead caused by the data structures.  Here we describe the details of the gate count optimal search. The same ideas were adopted for depth optimal search and can be extended to more general cost functions.  We did not target to study all possible optimizations in a systematic way. We present a set of solutions that allowed us to obtain the results in a reasonable amount of time and designed our software to be scalable enough to support different types of search. 

Possible costs of unitaries belong to a short range of integer values. Once we found all unitaries with some fixed cost we store them as a sorted array.  We call it a \emph{layer}. It allows us to quickly lookup unitaries with a given cost, however, it is expensive to ensure consistency of this data structure when inserting new elements into it. When searching for unitaries with specific cost we use C++ \emph{set} container to store only unique elements. We build layers one by one. To build the layer $k$ we pick an element of the layer $k-1$---we call it a \emph{parent} unitary.  Then we compose it with all possible gates and check if the resulting unitary was not found earlier.  The only possible costs of the resulting unitary are $k$, $k-1$, or $k-2$. If we get cost less than $k-2$ this contradicts the knowledge that the cost of the parent unitary is indeed $k-1$.  Therefore, during the search we need to keep only two previous layers in the memory.  We repeat the procedure for all unitaries in the layer $k-
1$.
  It can be executed in parallel for several parent unitaries. Only the addition of the unitaries with cost $k$ to the \emph{set} container must be synchronized. After the layer was built we copy content of the \emph{set} container into sorted array and start building a new layer. 


Finally, we describe how to find a circuit using the precomputed layers. If we find that a unitary belongs to the layer $k$ this means that there exists a circuit with $k$ gates computing the unitary. Therefore, by removing the last gate in the circuit we obtain an optimal circuit with $k-1$ gates which corresponds to a unitary with cost $k-1$.  By composing the source unitary with all possible gates and checking cost of the result we identify the last gate in the optimal circuit. We proceed further in a similar fashion, until we reach the canonical representative of the identity.  In the case when we rename inputs and output simultaneously we always get an identity in the end. When renaming of inputs and outputs is independent we obtain a circuit that is composed entirely of SWAP gates that represents a permutation of the inputs.

\section{Experimental results}
In this section we describe the results of our search together with the optimization experiments that rely on the databases of the optimal circuits we found. For the experiments that require more than 8 GB of RAM memory we used a high performance server with eight Quad-Core AMD Opteron 8356 (2.30 GHz) processors and 128 GB of RAM memory.  These are the experiments involving 4- and 5-qubit Clifford unitaries. For all other experiments we used a machine with a single quad-core Intel Core i7-2600 (3.40 GHz) processor and 8 GB of RAM.

\begin{table*}[t]
\begin{minipage}[t]{\columnwidth}%
\begin{tabular}{|c||c|c|c|c||c|c|}
\hline 
Code & $c_1$ & $c_{1o}$ & $c_2$ & $c_{2o}$ & $t_{1o}$ & $t_{2o}.$ \tabularnewline \hline \hline 
[[25,1,9]] & 440 & 285 & 387 & 205 & 22.2707 & 6.57012 \tabularnewline \hline \hline 
[[26,1,9]] & 444 & 287 & 389 & 207 & 22.8359 & 7.97804 \tabularnewline \hline 
[[26,4,8]] & 500 & 336 & 528 & 250 & 30.073 & 18.6791 \tabularnewline \hline \hline 
[[27,1,9]] & 592 & 396 & 479 & 241 & 31.8254 & 15.1547 \tabularnewline \hline 
[[27,2,9]] & 559 & 377 & 568 & 295 & 39.9342 & 18.5159 \tabularnewline \hline 
[[27,3,9]] & 566 & 373 & 566 & 274 & 38.629 & 10.6849 \tabularnewline \hline 
[[27,4,8]] & 504 & 335 & 530 & 252 & 28.0685 & 22.9666 \tabularnewline \hline 
[[27,8,6]] & 498 & 341 & 558 & 305 & 45.949 & 15.2595 \tabularnewline \hline 
[[27,9,6]] & 453 & 310 & 588 & 305 & 32.2922 & 17.1963 \tabularnewline \hline 
[[27,10,5]] & 428 & 293 & 563 & 293 & 59.7698 & 10.5993 \tabularnewline \hline 
[[27,11,5]] & 409 & 279 & 541 & 295 & 29.6078 & 12.5559 \tabularnewline \hline \hline 
[[28,0,10]] & 652 & 446 & 526 & 248 & 45.3604 & 18.2336 \tabularnewline \hline 
[[28,1,10]] & 660 & 446 & 531 & 284 & 41.0448 & 13.5861 \tabularnewline \hline 
[[28,2,10]] & 666 & 427 & 592 & 285 & 44.143 & 16.4625 \tabularnewline \hline 
[[28,3,9]] & 570 & 378 & 568 & 276 & 60.009 & 10.2351 \tabularnewline \hline \hline 
\end{tabular}
\end{minipage}
\begin{minipage}[t]{\columnwidth}%
\begin{tabular}{|c||c|c|c|c||c|c|}
\hline 
Code & $c_1$ & $c_{1o}$ & $c_2$ & $c_{2o}$ & $t_{1o}$ & $t_{2o}.$ \tabularnewline \hline \hline 
[[29,0,11]] & 726 & 479 & 597 & 288 & 63.5229 & 12.0342 \tabularnewline \hline 
[[29,1,11]] & 709 & 477 & 572 & 294 & 59.994 & 10.7589 \tabularnewline \hline 
[[29,2,10]] & 670 & 430 & 594 & 287 & 42.623 & 19.8844 \tabularnewline \hline 
[[29,3,9]] & 574 & 380 & 570 & 278 & 59.022 & 12.6315 \tabularnewline \hline 
[[29,4,8]] & 512 & 341 & 534 & 256 & 30.7405 & 30.2718 \tabularnewline \hline 
[[29,5,7]] & 492 & 305 & 518 & 263 & 29.5458 & 31.1485 \tabularnewline \hline 
[[29,6,7]] & 602 & 409 & 577 & 318 & 52.8024 & 14.8819 \tabularnewline \hline 
[[29,7,6]] & 549 & 376 & 593 & 298 & 28.9031 & 20.17 \tabularnewline \hline 
[[29,8,6]] & 488 & 318 & 576 & 313 & 45.6243 & 10.709 \tabularnewline \hline \hline 
[[30,0,12]] & 813 & 524 & 662 & 310 & 71.1554 & 18.601 \tabularnewline \hline 
[[30,1,11]] & 713 & 479 & 574 & 296 & 60.7773 & 12.9511 \tabularnewline \hline 
[[30,2,10]] & 674 & 432 & 596 & 289 & 39.7054 & 24.7658 \tabularnewline \hline 
[[30,4,8]] & 516 & 349 & 536 & 258 & 34.0143 & 34.5184 \tabularnewline \hline 
[[30,8,7]] & 627 & 425 & 707 & 378 & 75.6614 & 22.7352 \tabularnewline \hline \hline 
[[40,30,4]] & 452 & 311 & 679 & 362 & 198.226 & 41.9046 \tabularnewline \hline 
\end{tabular}
\end{minipage}
\caption[Code optimizations]
{\label{tab:qc}
The results of application of the peephole optimization to encoding circuits for Quantum Error Correcting codes. [[n,k,d]] denotes the code that uses n physical qubits, encodes k logical qubits and has distance d, $c_k$ -- number of gates in the circuit obtained using Algorithm $k$, $c_{ko}$ -- number of gates in the circuit after application of the peephole optimization using the database of 4-qubit optimal Clifford circuits, $t_{ko}$ -- runtime of peephole optimization software (in seconds, user time, using single core of Intel Core i7-2600) as applied to the circuits produced by the Algorithm $k$. 
}
\end{table*}

\subsection{Distribution of the optimal circuits}

We found optimal circuits for Clifford unitaries acting on 2 to 4 qubits (Figs.~\ref{fig:cl24}, \ref{fig:cl24-cz}) and optimal linear reversible circuits acting on up to 6 qubits.  In both cases we found both circuits with the optimal gate count and those with the minimal depth. For the case of Clifford unitaries we also found circuits with optimal number of controlled-Z gates (i.e., replacing the CNOT with the controlled-Z gate in the elementary gate set).

Distributions reported in Figs.~\ref{fig:cl24}, \ref{fig:cl24-cz} are interesting for the randomized benchmarking of quantum information processing systems. The benchmarking protocol \cite{rb} involves the application of a large number of randomly chosen Clifford unitaries.  Knowledge of the distribution of the number of gates allows us to estimate the average time required for each experiment, and evaluate its feasibility due to, e.g., the effects of the decoherence.  Using optimal circuits also minimizes the time required for the experiment.  Finally, this data may be used to estimate the average fidelity of the two-qubit gates used to perform the benchmarking protocol.  This is because it is based on the knowledge of the average number of two-qubit gates used \cite{rbmg}; latter follows directly from our results.

\subsection{Five-qubit Clifford unitaries\label{sec:5q}}

The search for five-qubit unitaries up to input/output order is challenging, but it is still tractable using modern computers.  The number of the different unitaries on five qubits is about $2.4 \times 10^{17}$ (Table~\ref{tab:sz}).  We need 100 bits to store each group element.  Factoring out simultaneous renaming of inputs and output allows us to reduce the size of the database by approximately 120 times.  However, one still needs at least $2.41 \times 10^6$ GB to store the full database in this case.  To allow the search of any 5-qubit Clifford unitary up to input/output order we allowed the independent renaming of the inputs and outputs of the circuits and used meet in the middle \cite{fbit} approach.  We synthesized all 5-qubit unitaries that use up to 11 gates which allowed us to search for unitaries that require up to 22 gates. It is unknown what is the maximum number of gates needed to implement any 5-qubit Clifford unitary.  We ran an experiment to estimate the distribution of the number of gates 
required to 
implement a unitary.  We used the algorithm described in \cite{rand} to generate uniformly distributed random Clifford unitaries and found their gate count.  The distribution of the number of gates for 5-qubit unitaries, shown in Fig.~\ref{fig:cl5}, was obtained using 20,000 samples.  We used Hoeffding inequality \cite{Hoeffding2012} to estimate errors for confidence level 0.999.  Based on the above calculation, we concluded that the use of the 11-layer database and the meet in the middle should allow finding optimal circuits for any 5-qubit Clifford unitary up to input/output order.

\begin{figure*}[t]
\begin{minipage}[t]{1.07\columnwidth}%
\includegraphics{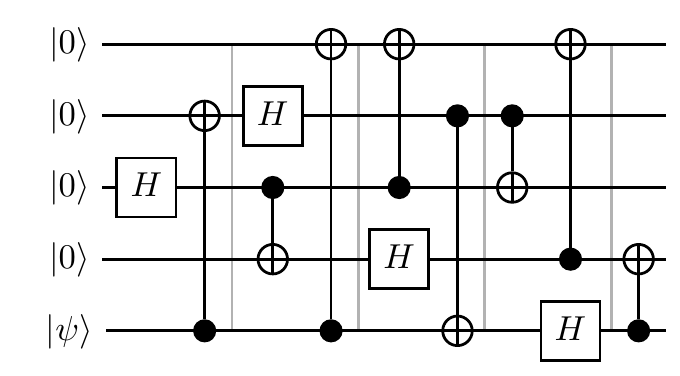}
\end{minipage}
\begin{minipage}[t]{0.93\columnwidth}
\includegraphics{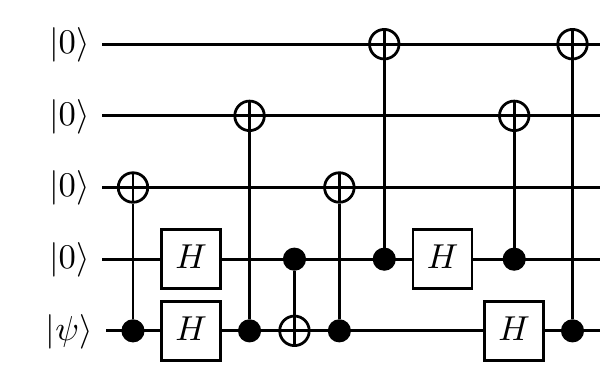}
\end{minipage}
\caption{\label{fig:5qcodeopt} Optimal encoding circuits for the five-qubit code: (left) depth optimal circuit, depth=5; (right) circuit with the minimal number of gates, being 11~gates.  Input marked $|\psi\rangle$ corresponds to the state being encoded.}
\end{figure*}
\begin{figure*}[t]
\begin{center}
\includegraphics{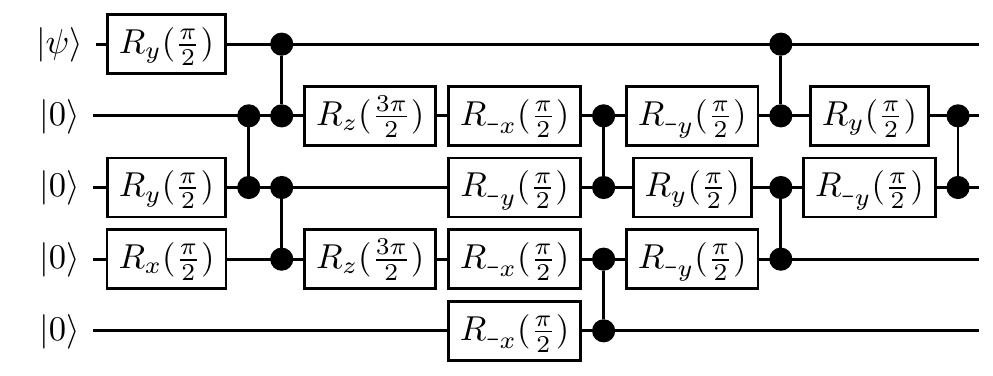}
\caption{\label{fig:5qcode} Encoding circuit for the five-qubit code used in \cite{5qexp}.  The two-qubit gate corresponds to $e^{-iZZ\pi/4}$.  Eight of them are required to implement the encoding circuit.}
\end{center}
\end{figure*}

\subsection{Peephole optimization} \label{subsec:peephole}

We used the database of the optimal 4-qubit Clifford circuits to perform peephole optimization. Briefly, peephole optimization works as follows \cite{pho}. First, choose a pivot gate from the circuit and enumerate all subcircuits including it and acting on the number of qubits less or equal to 4 (otherwise, some small parameter of choice, which in our case was 4).  Next, for each subcircuit find its cost and the optimal cost of the unitary that it computes. When beneficial, replace the less efficient subcircuit with its optimal version.  This procedure is repeated until it is no longer possible to reduce the cost of some subcircuit corresponding to some pivot element.  

When enumerating subcircuits we take it into account that some gates commute and we can build larger subcircuits by moving gates.  This requires us to examine the whole circuit at each step and results in quadratic complexity of the algorithm in the number of gates in the circuit.  In practice, the algorithm runtime depends on the circuit 
structure. Furthermore, for large size circuits a different and more efficient algorithm may be employed to find suitable subcircuits, including limiting the ``window'' in which the gates are to be found (limiting this window to a constant size results in the reduction of the algorithm complexity from quadratic to linear).  We did not investigate this further since in practice the subcircuit extraction algorithm with unbounded window did not take very long to complete for the circuits we tried.  A more detailed description of the peephole optimization may be found in \cite{pho}, along with a description of the version of the algorithm that produces slightly worse results, but requires linear time in the number of gates.

We applied peephole optimization to encoding circuits for quantum error correcting codes (QECCs).  To obtain an encoding circuit for QECC one starts with the stabilizer generators of the code and applies an algorithm that produces the encoding circuit.  We implemented two algorithms.  The first one is a version of the canonical decomposition theorem \cite{stab} for stabilizers that produces stages of CNOT, H, and P gates (Algorithm 1).  The second one~(Algorithm 2), taken from~\cite{gbook}, produces circuits that do not have an expressed staged structure.  Table~\ref{tab:qc} summarizes the results of our experiment with codes from \cite{codes}. The code for these experiments was not parallelized. Applying peephole optimization to the circuits produced by Algorithm 2 results in a reduction of the number of gates by 45-53\%. 

\subsection{Optimal encoding circuit for five-qubit quantum error correcting code}

Using a slightly modified version of our algorithm we found a depth optimal circuit for the five-qubit $[[5,1,3]]$ error correcting code.  This code encodes one qubit and corrects any single qubit error. In this case only first four out of 10 lines of the binary symplectic matrix are specified.  We first found depth optimal circuits that produce matrices with different first four lines. The problem has an extra degree of freedom---the addition of lines of the binary symplectic matrix to each other does not change the code.  In other words, left multiplication of the specified part of the binary symplectic matrix by $4\times 4$ invertible binary matrix leaves the code unchanged.  Search for all four-bit optimal linear reversible circuits gave us a database of all $4\times 4$ invertible binary matrices.  We used it to go through all matrices equivalent to the one that defines the five-qubit code.  Depth and gate count optimal circuits found are shown in Fig.~\ref{fig:5qcodeopt}. One of the best previously 
known circuits 
is 
illustrated in Fig.~\ref{fig:5qcode}.  Our approach may also be used to synthesize optimal circuits for other quantum error correcting codes that use a small number of qubits.


\section{Conclusions}
We explored the limitations of the brute force search for optimal circuits implementing Clifford and linear reversible unitaries.  Using typical memory and processing power available today, it is possible to search for up to four-qubit optimal Clifford unitaries and six-qubit linear reversible unitaries.  We also demonstrated that additional assumptions allow to search for optimal Clifford unitaries with up to five qubits. It is possible to make further assumptions resulting in greater sub-optimality, but reducing the size of the search space.  For example, one may allow to apply Hadamard gates to each output in the end of the circuit for free. This will further reduce the size of search space by approximately $2^n$, where $n$ is the number of qubits.  It is easy to come up with canonical form computation for this case.  Of course, circuits produced by the algorithm will not be exactly optimal. However, the results will be very close to optimal if the cost of Hadamard gates is small.  Using more restricted 
gate sets, 
such as those that allow only nearest neighbour or two nearest neighbour interactions has the opposite effect. In such case we do not have the symmetry between all qubits, which results in the growth of the search space. 

Using lookup in our database as a part of the peephole optimization shows that this is an efficient and promising approach for the optimization of larger Clifford circuits.


\section{Acknowledgements}
 Authors supported in part by the Intelligence Advanced Research Projects Activity (IARPA) via Department of Interior National Business Center Contract number DllPC20l66. The U.S. Government is authorized to reproduce and distribute reprints for Governmental purposes notwithstanding any copyright annotation thereon. Disclaimer: The views and conclusions contained herein are those of the authors and should not be interpreted as necessarily representing the official policies or endorsements, either expressed or implied, of IARPA, DoI/NBC or the U.S. Government.
 
 This material is based upon work partially supported by the National Science Foundation (NSF), during D. Maslov's assignment at the Foundation.
 Any opinion, findings, and conclusions or recommendations expressed in this material are those of the author(s) and do not necessarily reflect the views of the National Science Foundation.
 
 VK wishes to thank Michele Mosca for his helpful discussions.




\begin{thebibliography}{10}%
\makeatletter
\providecommand \@ifxundefined [1]{%
 \ifx #1\undefined \expandafter \@firstoftwo
 \else \expandafter \@secondoftwo
\fi
}%
\providecommand \@ifnum [1]{%
 \ifnum #1\expandafter \@firstoftwo
 \else \expandafter \@secondoftwo
\fi
}%
\providecommand \enquote [1]{``#1''}%
\providecommand \bibnamefont  [1]{#1}%
\providecommand \bibfnamefont [1]{#1}%
\providecommand \citenamefont [1]{#1}%
\providecommand\href[0]{\@sanitize\@href}%
\providecommand\@href[1]{\endgroup\@@startlink{#1}\endgroup\@@href}%
\providecommand\@@href[1]{#1\@@endlink}%
\providecommand \@sanitize [0]{\begingroup\catcode`\&12\catcode`\#12\relax}%
\@ifxundefined \pdfoutput {\@firstoftwo}{%
 \@ifnum{\z@=\pdfoutput}{\@firstoftwo}{\@secondoftwo}%
}{%
 \providecommand\@@startlink[1]{\leavevmode}%
 \providecommand\@@endlink[0]{}%
}{%
 \providecommand\@@startlink[1]{%
  \leavevmode
  \pdfstartlink
   attr{/Border[0 0 1 ]/H/I/C[0 1 1]}%
   user{/Subtype/Link/A<</Type/Action/S/URI/URI(#1)>>}%
  \relax
 }%
 \providecommand\@@endlink[0]{\pdfendlink}%
}%
\providecommand \url  [0]{\begingroup\@sanitize \@url }%
\providecommand \@url [1]{\endgroup\@href {#1}{\urlprefix}}%
\providecommand \urlprefix [0]{URL }%
\providecommand \Eprint[0]{\href }%
\@ifxundefined \urlstyle {%
  \providecommand \doi [1]{doi:\discretionary{}{}{}#1}%
}{%
  \providecommand \doi [0]{doi:\discretionary{}{}{}\begingroup
  \urlstyle{rm}\Url }%
}%
\providecommand \doibase [0]{http://dx.doi.org/}%
\providecommand \Doi[1]{\href{\doibase#1}}%
\providecommand \bibAnnote [3]{%
  \BibitemShut{#1}%
  \begin{quotation}\noindent
    \textsc{Key:}\ #2\\\textsc{Annotation:}\ #3%
  \end{quotation}%
}%
\providecommand \bibAnnoteFile [2]{%
  \IfFileExists{#2}{\bibAnnote {#1} {#2} {\input{#2}}}{}%
}%
\providecommand \typeout [0]{\immediate \write \m@ne }%
\providecommand \selectlanguage [0]{\@gobble}%
\providecommand \bibinfo [0]{\@secondoftwo}%
\providecommand \bibfield [0]{\@secondoftwo}%
\providecommand \translation [1]{[#1]}%
\providecommand \BibitemOpen[0]{}%
\providecommand \bibitemStop [0]{}%
\providecommand \bibitemNoStop [0]{.\EOS\space}%
\providecommand \EOS [0]{\spacefactor3000\relax}%
\providecommand \BibitemShut [1]{\csname bibitem#1\endcsname}%
\bibitem{arXiv:0707.0963}%
  \BibitemOpen
  \bibfield{author}{%
  \bibinfo {author} {\bibfnamefont{E.}~\bibnamefont{Knill}}, \bibinfo {author}
  {\bibfnamefont{D.}~\bibnamefont{Leibfried}}, \bibinfo {author}
  {\bibfnamefont{R.}~\bibnamefont{Reichle}}, \bibinfo {author}
  {\bibfnamefont{J.}~\bibnamefont{Britton}}, \bibinfo {author}
  {\bibfnamefont{R.~B.}\ \bibnamefont{Blakestad}}, \bibinfo {author}
  {\bibfnamefont{J.~D.}\ \bibnamefont{Jost}}, \bibinfo {author}
  {\bibfnamefont{C.}~\bibnamefont{Langer}}, \bibinfo {author}
  {\bibfnamefont{R.}~\bibnamefont{Ozeri}}, \bibinfo {author}
  {\bibfnamefont{S.}~\bibnamefont{Seidelin}},\ and\ \bibinfo {author}
  {\bibfnamefont{D.~J.}\ \bibnamefont{Wineland}},\ }%
  \bibfield{journal}{%
  \Doi{10.1103/PhysRevA.77.012307}{\bibinfo {journal} {Phys. Rev. A}}\ }%
  \textbf{\bibinfo {volume} {77}},\ \bibinfo {pages} {012307} (\bibinfo {month}
  {Jan}\ \bibinfo {year} {2008})%
  \bibAnnoteFile{NoStop}{arXiv:0707.0963}%
\bibitem{rbmg}%
  \BibitemOpen
  \bibfield{author}{%
  \bibinfo {author} {\bibfnamefont{J.~P.}\ \bibnamefont{Gaebler}}, \bibinfo
  {author} {\bibfnamefont{A.~M.}\ \bibnamefont{Meier}}, \bibinfo {author}
  {\bibfnamefont{T.~R.}\ \bibnamefont{Tan}}, \bibinfo {author}
  {\bibfnamefont{R.}~\bibnamefont{Bowler}}, \bibinfo {author}
  {\bibfnamefont{Y.}~\bibnamefont{Lin}}, \bibinfo {author}
  {\bibfnamefont{D.}~\bibnamefont{Hanneke}}, \bibinfo {author}
  {\bibfnamefont{J.~D.}\ \bibnamefont{Jost}}, \bibinfo {author}
  {\bibfnamefont{J.~P.}\ \bibnamefont{Home}}, \bibinfo {author}
  {\bibfnamefont{E.}~\bibnamefont{Knill}}, \bibinfo {author}
  {\bibfnamefont{D.}~\bibnamefont{Leibfried}},\ and\ \bibinfo {author}
  {\bibfnamefont{D.~J.}\ \bibnamefont{Wineland}},\ }%
  \bibfield{journal}{%
  \Doi{10.1103/PhysRevLett.108.260503}{\bibinfo {journal} {Phys. Rev. Lett.}}\
  }%
  \textbf{\bibinfo {volume} {108}},\ \bibinfo {pages} {260503} (\bibinfo
  {month} {Jun}\ \bibinfo {year} {2012})%
  \bibAnnoteFile{NoStop}{rbmg}%
\bibitem{rb2q}%
  \BibitemOpen
  \bibfield{author}{%
  \bibinfo {author} {\bibfnamefont{A.~D.}\ \bibnamefont{C\'orcoles}}, \bibinfo
  {author} {\bibfnamefont{J.~M.}\ \bibnamefont{Gambetta}}, \bibinfo {author}
  {\bibfnamefont{J.~M.}\ \bibnamefont{Chow}}, \bibinfo {author}
  {\bibfnamefont{J.~A.}\ \bibnamefont{Smolin}}, \bibinfo {author}
  {\bibfnamefont{M.}~\bibnamefont{Ware}}, \bibinfo {author}
  {\bibfnamefont{J.}~\bibnamefont{Strand}}, \bibinfo {author}
  {\bibfnamefont{B.~L.~T.}\ \bibnamefont{Plourde}},\ and\ \bibinfo {author}
  {\bibfnamefont{M.}~\bibnamefont{Steffen}},\ }%
  \bibfield{journal}{%
  \Doi{10.1103/PhysRevA.87.030301}{\bibinfo {journal} {Phys. Rev. A}}\ }%
  \textbf{\bibinfo {volume} {87}},\ \bibinfo {pages} {030301} (\bibinfo {month}
  {Mar}\ \bibinfo {year} {2013})%
  \bibAnnoteFile{NoStop}{rb2q}%
\bibitem{arXiv:0808.3973}%
  \BibitemOpen
  \bibfield{author}{%
  \bibinfo {author} {\bibfnamefont{C.~A.}\ \bibnamefont{Ryan}}, \bibinfo
  {author} {\bibfnamefont{M.}~\bibnamefont{Laforest}},\ and\ \bibinfo {author}
  {\bibfnamefont{R.}~\bibnamefont{Laflamme}},\ }%
  \bibfield{journal}{%
  \Doi{10.1088/1367-2630/11/1/013034}{\bibinfo {journal} {New Journal of
  Physics}}\ }%
  \textbf{\bibinfo {volume} {11}},\ \bibinfo {pages} {013034} (\bibinfo {month}
  {Jan}\ \bibinfo {year} {2009})%
  \bibAnnoteFile{NoStop}{arXiv:0808.3973}%
\bibitem{pho}%
  \BibitemOpen
  \bibfield{author}{%
  \bibinfo {author} {\bibfnamefont{A.~K.}\ \bibnamefont{Prasad}}, \bibinfo
  {author} {\bibfnamefont{V.~V.}\ \bibnamefont{Shende}}, \bibinfo {author}
  {\bibfnamefont{I.~L.}\ \bibnamefont{Markov}}, \bibinfo {author}
  {\bibfnamefont{J.~P.}\ \bibnamefont{Hayes}},\ and\ \bibinfo {author}
  {\bibfnamefont{K.~N.}\ \bibnamefont{Patel}},\ }%
  \bibfield{journal}{%
  \Doi{10.1145/1216396.1216399}{\bibinfo {journal} {ACM Journal on Emerging
  Technologies in Computing Systems}}\ }%
  \textbf{\bibinfo {volume} {2}},\ \bibinfo {pages} {277} (\bibinfo {month}
  {Oct.}\ \bibinfo {year} {2006})%
  \bibAnnoteFile{NoStop}{pho}%
\bibitem{stab}%
  \BibitemOpen
  \bibfield{author}{%
  \bibinfo {author} {\bibfnamefont{S.}~\bibnamefont{Aaronson}}\ and\ \bibinfo
  {author} {\bibfnamefont{D.}~\bibnamefont{Gottesman}},\ }%
  \bibfield{journal}{%
  \Doi{10.1103/PhysRevA.70.052328}{\bibinfo {journal} {Phys. Rev. A}}\ }%
  \textbf{\bibinfo {volume} {70}},\ \bibinfo {pages} {052328} (\bibinfo {month}
  {Nov.}\ \bibinfo {year} {2004})%
  \bibAnnoteFile{NoStop}{stab}%
\bibitem{quant-ph/0703211}%
  \BibitemOpen
  \bibfield{author}{%
  \bibinfo {author} {\bibfnamefont{D.}~\bibnamefont{Maslov}},\ }%
  \bibfield{journal}{%
  \Doi{10.1103/PhysRevA.76.052310}{\bibinfo {journal} {Phys. Rev. A}}\ }%
  \textbf{\bibinfo {volume} {76}},\ \bibinfo {pages} {052310} (\bibinfo {month}
  {Nov}\ \bibinfo {year} {2007})%
  \bibAnnoteFile{NoStop}{quant-ph/0703211}%
\bibitem{hrepr}%
  \BibitemOpen
  \bibfield{author}{%
  \bibinfo {author} {\bibfnamefont{D.}~\bibnamefont{Gottesman}},\ }%
  in\ \emph{\bibinfo {booktitle} {Proc. of the XXII International Colloquium on
  Group Theoretical Methods in Physics}}\ (\bibinfo {publisher} {International
  Press},\ \bibinfo {address} {Cambridge, MA},\ \bibinfo {year} {1999})\ pp.\
  \bibinfo {pages} {32--43},\
  \Eprint{http://arxiv.org/abs/quant-ph/9807006}{arXiv:quant-ph/9807006}%
  \bibAnnoteFile{NoStop}{hrepr}%
\bibitem{fbit}%
  \BibitemOpen
  \bibfield{author}{%
  \bibinfo {author} {\bibfnamefont{O.}~\bibnamefont{Golubitsky}}\ and\ \bibinfo
  {author} {\bibfnamefont{D.}~\bibnamefont{Maslov}},\ }%
  \bibfield{journal}{%
  \Doi{10.1109/TC.2011.144}{\bibinfo {journal} {IEEE Transactions on
  Computers}}\ }%
  \textbf{\bibinfo {volume} {61}},\ \bibinfo {pages} {1341} (\bibinfo {month}
  {Sep.}\ \bibinfo {year} {2012})%
  \bibAnnoteFile{NoStop}{fbit}%
\bibitem{Note1}%
  \BibitemOpen
  \bibinfo {note} {By row permutation we mean a permutation acting
  simultaneously on first $n$ rows and rows $n+1,\protect \ldots , 2n$.}%
  \bibAnnoteFile{Stop}{Note1}%
\bibitem{rb}%
  \BibitemOpen
  \bibfield{author}{%
  \bibinfo {author} {\bibfnamefont{E.}~\bibnamefont{Magesan}}, \bibinfo
  {author} {\bibfnamefont{J.}~\bibnamefont{Gambetta}},\ and\ \bibinfo {author}
  {\bibfnamefont{J.}~\bibnamefont{Emerson}},\ }%
  \bibfield{journal}{%
  \Doi{10.1103/PhysRevA.85.042311}{\bibinfo {journal} {Phys. Rev. A}}\ }%
  \textbf{\bibinfo {volume} {85}},\ \bibinfo {pages} {042311} (\bibinfo {month}
  {Apr.}\ \bibinfo {year} {2012})%
  \bibAnnoteFile{NoStop}{rb}%
\bibitem{rand}%
  \BibitemOpen
  \bibfield{author}{%
  \bibinfo {author} {\bibfnamefont{D.}~\bibnamefont{DiVincenzo}}, \bibinfo
  {author} {\bibfnamefont{D.}~\bibnamefont{Leung}},\ and\ \bibinfo {author}
  {\bibfnamefont{B.}~\bibnamefont{Terhal}},\ }%
  \bibfield{journal}{%
  \Doi{10.1109/18.985948}{\bibinfo {journal} {IEEE Transactions on Information
  Theory}}\ }%
  \textbf{\bibinfo {volume} {48}},\ \bibinfo {pages} {580} (\bibinfo {month}
  {Mar.}\ \bibinfo {year} {2002})%
  \bibAnnoteFile{NoStop}{rand}%
\bibitem{Hoeffding2012}%
  \BibitemOpen
  \bibfield{author}{%
  \bibinfo {author} {\bibfnamefont{W.}~\bibnamefont{Hoeffding}},\ }%
  \bibfield{journal}{%
  \bibinfo {journal} {Journal of the American Statistical Association}\ }%
  \textbf{\bibinfo {volume} {58}},\ \bibinfo {pages} {13} (\bibinfo {month}
  {Mar.}\ \bibinfo {year} {1963})%
  \bibAnnoteFile{NoStop}{Hoeffding2012}%
\bibitem{5qexp}%
  \BibitemOpen
  \bibfield{author}{%
  \bibinfo {author} {\bibfnamefont{E.}~\bibnamefont{Knill}}, \bibinfo {author}
  {\bibfnamefont{R.}~\bibnamefont{Laflamme}}, \bibinfo {author}
  {\bibfnamefont{R.}~\bibnamefont{Martinez}},\ and\ \bibinfo {author}
  {\bibfnamefont{C.}~\bibnamefont{Negrevergne}},\ }%
  \bibfield{journal}{%
  \Doi{10.1103/PhysRevLett.86.5811}{\bibinfo {journal} {Phys. Rev. Lett.}}\ }%
  \textbf{\bibinfo {volume} {86}},\ \bibinfo {pages} {5811} (\bibinfo {month}
  {Jun.}\ \bibinfo {year} {2001})%
  \bibAnnoteFile{NoStop}{5qexp}%
\bibitem{gbook}%
  \BibitemOpen
  \bibfield{author}{%
  \bibinfo {author} {\bibfnamefont{D.}~\bibnamefont{Gottesman}},\ }%
  \enquote{\bibinfo {title} {{(Unpublished) Lecture notes for QIC 890,
  University of Waterloo}},}\  (\bibinfo {year} {2012})%
  \bibAnnoteFile{NoStop}{gbook}%
\bibitem{codes}%
  \BibitemOpen
  \bibfield{author}{%
  \bibinfo {author} {\bibfnamefont{M.}~\bibnamefont{Grassl}},\ }%
  \enquote{\bibinfo {title} {{Encoding Circuits for Quantum Error-Correcting
  Codes (last accessed April 8, 2013)}},}\  (\bibinfo {year} {2007}),\
  \url{http://i20smtp.ira.uka.de/home/grassl/QECC/circuits/index.html}%
  \bibAnnoteFile{NoStop}{codes}%
\end{thebibliography}

%

\end{document}